\documentclass[journal,onecolumn]{IEEEtran}
\usepackage{hyperref}
\usepackage{graphicx}
\usepackage{color}
\usepackage{dcolumn}
\usepackage{bm}
\usepackage[latin1]{inputenc}
\usepackage{footnote}
\usepackage{epstopdf}
\usepackage{cite}
\usepackage{url}
\usepackage{amsmath}
\usepackage{amssymb}
\usepackage{balance}

\def \micron {$\mu$m~}

\begin{document}
\title{3D FEM Modelling of CORC\textsuperscript{\textregistered} Commercial Cables with Bean's like magnetization currents and its AC-Losses Behaviour}

\author{M.~U.~Fareed, M.~Kapolka, B.~C.~Robert, M.~Clegg, and H. S. Ruiz
\thanks{Manuscript Accepted at IEEE Transactions on Applied Superconductivity 18 January 2022. Submitted 20 September 2021.}
\thanks{The authors are with the College of Science and Engineering $\&$ Space Park Leicester, University of Leicester, Leicester LE1 7RH, United Kingdom
(e-mail: \href{mailto:dr.harold.ruiz@leicester.ac.uk}{dr.harold.ruiz@leicester.ac.uk})}
\thanks{The authors acknowledge the use of the High-Performance Computing Cluster Facilities (ALICE) provided by the University of Leicester.  M.U.F and M.C thanks the CSE and EPSRC-DTP studentships provided by the University of Leicester. M.U.F and M.K. shared first authorship. This work was supported by the UK Research and Innovation, Engineering and Physical Sciences Research Council (EPSRC), through the grant Ref. EP/S025707/1 led by H.S.R.}
}


%
\maketitle
\begin{abstract}
The Conductor on Rounded Core (CORC\textsuperscript{\textregistered}) cables manufactured by Advanced Conductor Technologies with current densities beyond 300 Amm$^{-2}$ at 4.2 K, and bending diameter of up to 3.5~cm, are considered as one of the strongest candidates for the next generation of high field power applications and magnets. In this paper, we present a full 3D FEM model for their monolayer and bilayer CORC\textsuperscript{\textregistered} cables made with up to three and six superconducting tapes respectively, disclosing the full curve of AC losses for the monolayer cable at magnetic fields beyond 60 mT, and the actual distribution of current density along and across the thickness of the superconducting tapes in both designs. The model is based on the so-called H-formulation, allowing to incorporate the true three-dimensionality of the tapes without recurring to 2D thin-film approaches where non-physical surface currents that do not follow the celebrated Bean's model for type-II superconductors appear. Likewise, good agreement with the experimentally measured AC-losses for the monolayer and bilayer cable have been obtained, with all the details of the model disclosed in this paper. 

\end{abstract}

\IEEEpeerreviewmaketitle


\section{Introduction}\label{Sec.1}

\IEEEPARstart{T}{he} development of High Temperature Superconducting (HTS) applications for ultra-lightweight generators, motors, or high field magnets, demand of a deep understanding of the convoluted performance of superconducting cables either for power connections, transmission lines, or for their large variety of coil windings, opening routes for multiple designs in the competition for the best high power superconducting cable. Nowadays, the default HTS tapes used for high current capacity and high magnetic field applications is the set of the so-called second generation (2G) coated conductors available in the market, whose theoretical limit for the upper critical field $B_{c2}$ is above 100 T \cite{Larbalestier2001}. 

Over time there has been three major cabling concepts put forward, the ROEBEL cable \cite{Goldacker2007RACC,Goldacker2014}, the Twisted Stacked Tape Cable (TSTC) \cite{Takayasu2012,Takayasu2017} and the Conductor on Round Core (CORC\textsuperscript{\textregistered}) cable \cite{vanderLaan2009SuST,vanderLaan2011}. In brief, in the TSTC concept, the cable contains several coated conductors stacked on top of each other, after which the bundle is twisted for improving bending performance. On the other hand, the ROEBEL cable is fairly inflexible when bent in plane of the stacked coated conductors, limiting therefore the range of magnets suitable for this kind of cables. Finally, due to the simplicity and low cost, the CORC\textsuperscript{\textregistered} cable represents the most attractive option regarding to the bending features, as by wounding the 2G-HTS tapes around an elastic rounded-former of e.g., copper or stainless steel, a full transposition of the tapes along the cable length can be obtained\cite{Kovacs2019,Souc2017SUST}. In fact, it has been already proved that CORC\textsuperscript{\textregistered} cables can be wounded in a small diameter coil with high magnetic fields up to 20 T \cite{Mulder2018IEEE,vanderLaan2019SUST}. 

Additionally, for the development of high-power superconducting cables, there is another requirement to consider which is attaining very low AC-losses at high transport currents and high magnetic fields. This is because significant hysteresis losses are caused when either the self or any external magnetic field is perpendicular to the wide surface of the HTS tapes that compose the cable. This is partially solved by the tapes' transposition in the CORC\textsuperscript{\textregistered} cable~\cite{Cobb2002,Souc2013}, with further solutions such as striating the tapes for a higher reduction on their AC-losses~\cite{Vojenciak2015SUST}. However, the designing of multilayer CORC\textsuperscript{\textregistered} cables with electric contact between the  tapes, generate coupling currents and hence adds more AC losses. Still, current sharing between the tapes is essential for cable stability with the inter-tape contact resistive parameter. All of this make the CORC\textsuperscript{\textregistered} cable geometry a complex structure without possible analytical solution for the calculation of AC losses, and for which topological reductions from the actual 3D geometry into 2D abstractions of the coated conductors (or the entire cable), can result in rather complex theoretical predictions with doubtful distributions of current density which do not seem to follow the celebrated Bean's model for type-II superconductors~\cite{QuanLi2014EoMC,FukuiS2006AoAL,JiangZhenan2008,Stenvall2013SUST,Wang2019SUST}. This is precisely the case in all previous attempts to model the full tree-dimensionality of CORC\textsuperscript{\textregistered} cables with the acclaimed H-formulation~\cite{Sheng2017IEEE,Tian2021IEEEa,Tian2021IEEEb,Yang2021JSNM}, as either the minimum length of the finite elements across the 1 micron thickness of the superconducting film seem to match the thickness of the renormalized layer or, not sufficient elements have been defined across this dimension to allow the actual penetration of magnetization currents. 

From the physical point of view, for a correct computation of the electromagnetic properties inside a superconducting film with an aspect ratio as high as the one of coated conductors, i.e., with a SC layer of just 1 micron thickness by 4-12 mm width, it is not sufficient to just re-dimension the thickness of the SC layer by a factor generally greater than 10, and then to renormalise its corresponding critical current density but, sufficient room (number of finite elements) should be enabled across all its dimensions to allow for an adequate solution of the Ampere's law. In this sense, if across the thickness of the SC layer no more than one 'row' of finite elements can be identified~\cite{Sheng2017IEEE,Tian2021IEEEa,Tian2021IEEEb,Yang2021JSNM}, then loops of current density such as magnetization currents might not close across the thickness of the superconductor, it due to the lack of a sufficiently large number of finite elements along this dimension. In consequence, the common and well-known occurrence of opposite profiles of current density (magnetization currents) across the thickness of a superconducting material, which is characteristic for all known type-II superconductors, have not been seen or disclosed in any of the previous 3D numerical approaches. Being this the main difference with our numerical approach, it is worth reminding that from the macroscopic point of view and the mathematical solution of the PDE system under a reduced scheme of finite elements, there is no reason why global averaged quantities such as the AC losses or the overall sample magnetization could not be accurately calculated by any of the approaches mentioned above. This even though no physical meaning could be given to the distribution of current density at a local level (inside the superconductor). Moreover, it results remarkable to find the existence of a further gap in the literature, as none of the previous numerical approaches attempted to reproduce the experimentally measured AC losses for the CORC\textsuperscript{\textregistered} cable originally introduced by Advanced Conductor Technologies~\cite{Majoros2014}.

\begin{figure}
\begin{center}
{\includegraphics[width=0.9\textwidth]{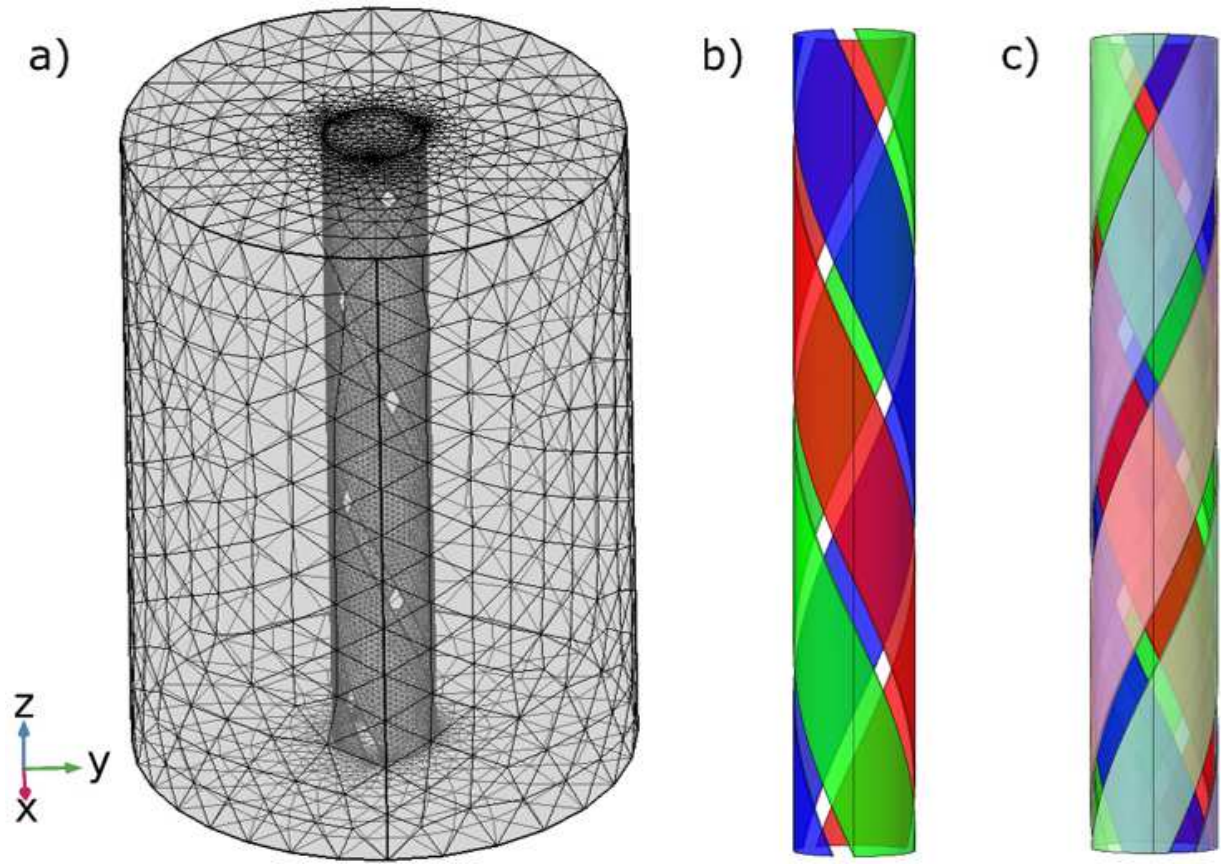}}
\caption{\label{Fig_1}Illustration of the full 3D CAD model for the CORC\textsuperscript{\textregistered} cable visualizing a) the meshed domain around the HTS tapes, b) the helical structure of the 1-layer cable with 3 tapes and, c) the 2-layers cable with a total of 6 tapes wounded in opposite direction between the layers. The outer layer of tapes is shadowed for better visibility of the tapes in the inner layer, and the length of both cables are not shown in the same scale as can be read from Table~\ref{Table_1}.
}
\end{center}
\end{figure}

Then, since there is no direct analytical solution for the calculation of AC losses in the CORC\textsuperscript{\textregistered} cable structure, various numerical simplifications have been utilised to tackle this problem. The most basic approaches assume 2D cross-sectional models where the helical structure of the cable is neglected, or extrinsically introduced by an empirical modification of the critical current density of the 2G-HTS tapes that compose the cable~\cite{QuanLi2014EoMC,FukuiS2006AoAL,JiangZhenan2008}. However, these approaches neglect the hysteresis losses caused by the twist pitch angle, i.e., by the non-perpendicularity of the derived magnetic field and the local profiles of current density along the entire length of tapes and cable, hence leading to unreliable results in terms of the AC losses specially at short twist pitch lengths. However, as mentioned above this difference can be overcome within a 2D model, by empirically adding the self-field hysteresis loss caused by the spiral structure of the wound tapes~\cite{Stenvall2013SUST}. Albeit, this method neglects the thickness of the HTS tapes and assume an infinitely thin film structure for the superconducting tapes. Then, to consider the full three-dimensional structure of the helicoidal arrangement of realistic HTS tapes in CORC\textsuperscript{\textregistered} cables, a full 3D model needs to be deployed in a formulation such as the $H$-formulation~\cite{Sheng2017IEEE,Wang2019SUST,Tian2021IEEEa,Tian2021IEEEb,Yang2021JSNM,RUIZ2021CEC-CORC}. Other formulations such as the $T-A$ formulation~\cite{Wang2019SUST} allows to consider the helicity of the CORC\textsuperscript{\textregistered} cable in 3D, but ultimately the dimensions of the HTS tapes are reduced to a 2D approach by assuming these as infinitely thin superconducting films. By this reason, from the electromagnetic point of view the $T-A$ approach cannot be considered as a fully 3D electromagnetic model, as any physical property inside of the superconducting tapes such as the local distribution of current density cannot be reliably calculated. Still, macroscopically averaged quantities such as the magnetization and AC losses of the CORC\textsuperscript{\textregistered}  cable can be obtained within this approach.

\begin{center}
\begin{table}[t]
\centering
\caption{\label{Table_1} Parameters of the modelled CORC cables.} 
\begin{tabular}{@{}l*{10}{l}}
\hline
Sample & FID$^{\ast}$ & FOD$^{\dag}$ & TW$^{\ddagger}$ & NTL$^{\ast\ast}$ & NLC$^{\dag\dag}$ & CTP$^{\ddagger\ddagger}$\\
~~\cite{Majoros2014} & (mm) & (mm) & (mm) &  &  & (mm) \\
\hline
~~B  & 3.13 & 4.76 & 4 & 3 & 1 & 40\\
~~C & 3.13 & 4.76 & 4 & 3 & 2 & 34\\
\hline
\end{tabular}\\ \vspace*{0.1cm}
$^{\ast}$ Former internal diameter, 
$^{\dag}$ Former outer diameter, 
$^{\ddagger}$ Tape width, 
$^{\ast\ast}$ Number of tapes per layer, 
$^{\dag\dag}$ Number of layers in cable, 
$^{\ddagger\ddagger}$ Cable Twist Pitch with alternating twist direction in layers for sample C.
\end{table}
\end{center}

Thus, in a fully 3D H-formulation revealing the actual distribution of current density within the finite thickness of the HTS tapes, and the derived AC losses, in this paper we present a detailed analysis of the one and two-layer CORC\textsuperscript{\textregistered} cables with three tapes per layer (see Fig.~\ref{Fig_1}), corresponding to the samples B and C in~\cite{Majoros2014} for which their main parameters are summarized in Table~\ref{Table_1}. For reproducibility of our results and the benefit of our readers, the implemented $H$-formulation in the general PDE framework of  COMSOL Multiphysics is explained in section $\ref{Sec.2}$, with particular mention to the quality of finite elements required for the occurrence of Bean's like magnetization currents across the thickness of the superconducting layers. This has been demonstrated in the case of the three SC4050 monolayer and bilayer open-ends CORC\textsuperscript{\textregistered} cables reported by Advanced Conductor Technologies in~\cite{Majoros2014}, which contains the experimental measurements of their AC losses under a 50~Hz magnetic field applied perpendicularly to its longitudinal axis. To ensure that the magnetization currents response of the CORC\textsuperscript{\textregistered} cable follows the distinctive features of Bean's model valid for any type-II superconductor, we present the full 3D current loops inside the HTS tapes in Sec.~\ref{Sec.3}. In the same section, we show the AC loss analysis with direct comparisons between the measured samples and our computational results, disclosing the actual AC losses of the 1-layer CORC\textsuperscript{\textregistered} cable at high magnetic fields, where the reported experimental measurements were affected by an unintended heating effect. On the other hand, for the 2-layer CORC\textsuperscript{\textregistered} cable our numerical results show a good agreement with the experimental measurements, showing the potential of this model to predict the electromagnetic response of CORC\textsuperscript{\textregistered} cables with many more tapes or layers. Finally, Sec.~\ref{Sec.4} is devoted to summarize the main conclusions of this study.

\section{Computational model and sample geometry}
\label{Sec.2}

The 3D electromagnetic modelling of CORC\textsuperscript{\textregistered} cables in the $H$-formulation has been implemented by the Partial Differential Equation (PDE) solver provided by COMSOL Multiphysics, it mainly due to its large popularity in the community of applied superconductivity but mostly due to its easy adaptation of diverse Finite Element Methods (FEM). The state variables of the $H$-formulation or Magnetic Field are commonly well known. Therefore, we will focus on their implementation inside the PDE module, where the general PDE form for the COMSOL interface is stated as:   
\begin{eqnarray}\label{Eq_1}
e_{a}\frac{\partial ^{2}{\bf H}}{\partial t^{2}}+ 
d_{a}\frac{\partial {\bf H}}{\partial t}+
\nabla \cdot \Gamma = f \,,
\end{eqnarray}
where $e_{a}$ is called the mass coefficient, $d_{a}$ is the damping coefficient, $\Gamma$ plays the role of a conservative vectorial flux,  and the function \textit{f} allows to define any source terms. The state variables of the magnetic field in Cartesian coordinates are represented as the matrix vector $\bf{H}=[H_{x}, H_{y}, H_{z}]$, and hence $\Gamma$ in the equation~\ref{Eq_1} must be transformed to resemble the left side of Faraday's law $\nabla\times\textbf{E}=-\mu_{0}\partial_{t}\textbf{H}$, in the same matrix notation. 

For this reason, starting with the divergence of the conservative flux vector $\Gamma = [\Gamma_{x}, \Gamma_{y}, \Gamma_{z}]$, which in the most general way implies the existence of non-isotropic components with space-dependent derivatives, i.e., 
\begin{eqnarray}
\label{Eq_2}
\nabla \cdot \Gamma =\left[ \begin{array}{ccc} 
\partial _{x}\Gamma_{11}+\partial _{y}\Gamma_{12}+\partial _{z}\Gamma_{13} \\
\partial _{x}\Gamma_{21}+\partial _{y}\Gamma_{22}+\partial _{z}\Gamma_{23} \\
 \partial _{x}\Gamma_{31}+\partial _{y}\Gamma_{32}+\partial _{z}\Gamma_{33}
 \end{array} \right]
\, ,
\end{eqnarray}
it is possible to infer between the left side of Faraday law and $\Gamma$, that the conservative flux vector results defined by  
\begin{eqnarray}\label{Eq_3}
\Gamma =\left[ \begin{array}{ccc} 
\Gamma_{11} & \Gamma_{12} & \Gamma_{13} \\
\Gamma_{21} & \Gamma_{22} & \Gamma_{23} \\
\Gamma_{31} & \Gamma_{32} & \Gamma_{33}
 \end{array} \right]=\left[ \begin{array}{ccc} 
0 & E_{z} & -E_{y} \\
-E_{z} & 0 & E_{x} \\
E_{y} & -E_{x} & 0
\end{array} \right]
\, .
\end{eqnarray}

Likewise, by assuming the mass coefficient $e_{a}=0$, the second time derivative in Eq.~\ref{Eq_1} is neglected, leaving only the damping term $d_{a}\partial_{t}\bf{H}$ which resembles the right side of Faraday's law by $\mu_{0}\partial_{t}\bf{H}$ with  
\begin{eqnarray}\label{Eq_4}
d_a =\left[ \begin{array}{ccc} 
da_{11} & da_{12} & da_{13} \\
da_{21} & da_{22} & da_{23} \\
da_{31} & da_{32} & da_{33}
 \end{array} \right]=\left[ \begin{array}{c c c} 
\mu_0 & 0 & 0\\
0 & \mu_0 & 0\\
0 & 0 & \mu_0
 \end{array} \right]
\, .
\end{eqnarray}
%

\begin{figure}
\begin{center}
{\includegraphics[width=0.9\textwidth]{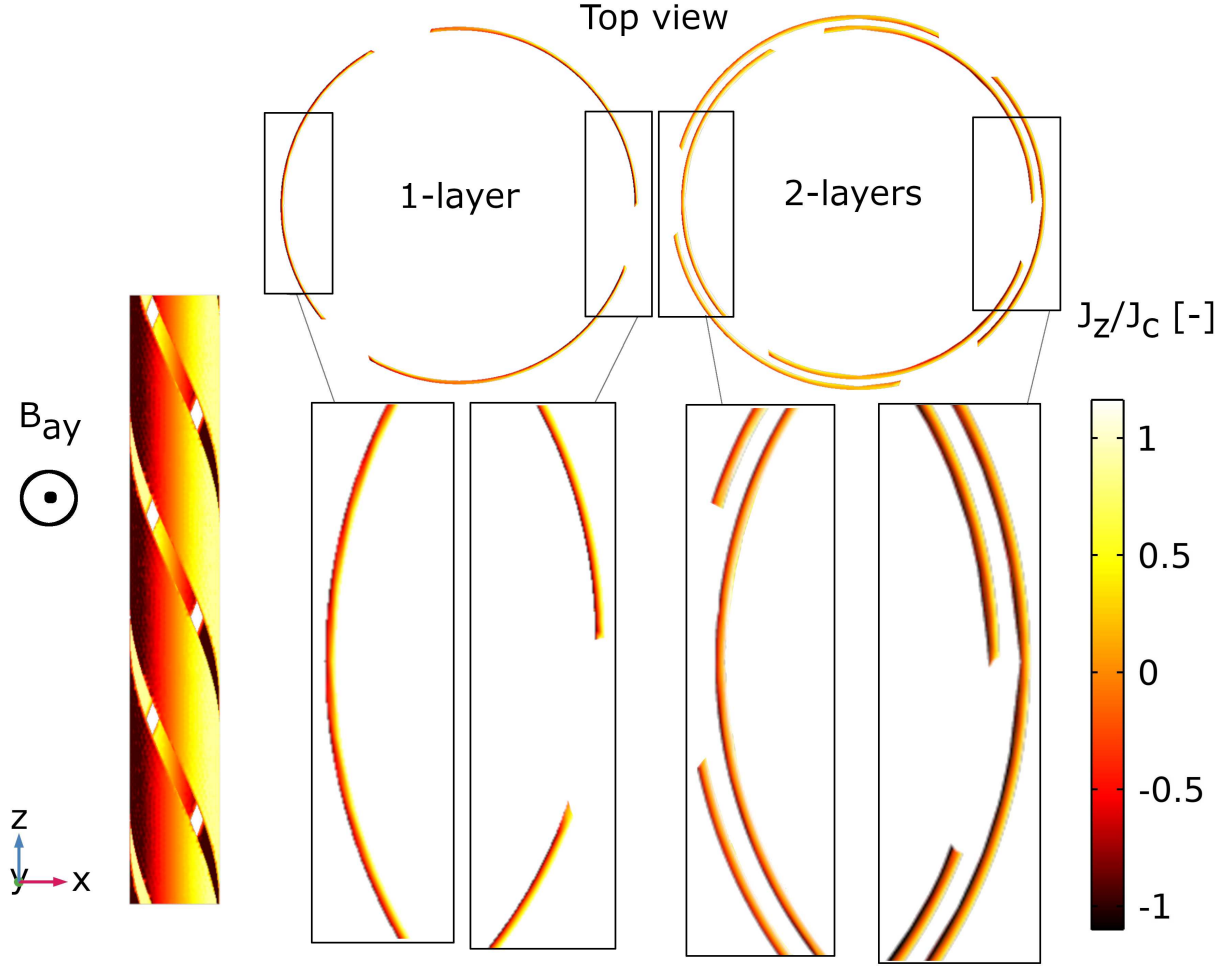}}
\caption{\label{Fig_2} At the left, the $J_{z}$ and main component of the magnetization currents at a 20 mT peak applied magnetic field in the monolayer CORC\textsuperscript{\textregistered} cable is shown. For better visualization a top view in the $XY$ plane of the monolayer and bilayer cables is also displayed. It shows magnetization currents closing along the tilted width of the helically wound tape, i.e., with positive and negative sign at the inner and outer side of the HTS tape, respectively.
}
\end{center}
\end{figure}

Then, for modelling the mono and bi-layer CORC\textsuperscript{\textregistered} cables manufactured by Advanced Conductor Technologies and characterized at the Ohio State University by Majoros et al., \cite{Majoros2014}, it is necessary to describe its general structure. The mono layer cable contains 3 tapes as seen in Fig.~\ref{Fig_1}~b, each made by the 4~mm wide SuperPower tape (SCS4050) with a critical current of 25 A per mm width, measured at self-field and LN2 temperature (77~K).  The bi-layer version contains 3 tapes per layer (Fig.~\ref{Fig_1} c) with each layer wound in counter direction. Cable dimensions and twist pitch measurements can be found in Table~\ref{Table_1}, where all CORC\textsuperscript{\textregistered} cables were measured with open ends~\cite{Majoros2014} allowing a straightforward representation of the cables within the 3D computational model. The actual thickness of the SCS4050 is of $\sim 100~\mu$m, including a $\sim1.8$~\micron Ag top-coating layer and a 20~\micron Cu stabilizer layer, the latter coating the YBCO tape which is deposited on a $\sim0.2$~\micron proprietary buffer over a $\sim50$~\micron Hastelloy C-276 substrate. As neither the metallic, buffer, or substrate layers have a magnetic signal nor are expected to carry electrical currents under transport current conditions below $I_{c}$, these layers can be modelled with the same physical properties of  what in the jargon for electromagnetic modelling is called as the ``air'' domain, i.e., a no magnetic highly resistive media. However, for  reducing the computing time and meshing requirements of the models, the thickness of the YBCO layer in the SCS4050 tape has been increased from $1~\mu$m to $50~\mu$m, renormalizing the critical current density accordingly, but ensuring the YBCO tape is placed in the middle of the $\sim 100~\mu$m thickness reserved for the SCS405 tape, such that its impact on the calculation of the H-field around the superconducting tapes is minimized.

From the physical point of view, a further caution must be taken in what concerns to the minimum length of the finite elements used for meshing the superconducting domain. In fact, if the mesh is not sufficiently fine and the minimum length of the devised finite elements is not less than the thickness of the renormalized layer (i.e., $50~\mu$m), then the solution of Ampere's law will be artificially forced in such way that proper Bean's like loops of magnetization current would not be able to appear. Nevertheless, by significantly reducing the minimum length of the finite elements, e.g., to clearly fit more than one element across the thickness of the SC layer, this can easily render to the same computational issues derived from the aspect-ratio problem for which the thickness of the SC layer was initially renormalized. In this sense, finite elements with minimum lengths close to the thickness of the renormalized layer can be used as long the higher degree of shape function (Lagrange elements) of at least quadratic order are used~\cite{FEM_Bathe_2006}, otherwise by using reduced finite element approaches such as the curl-elements, these could also limit the occurrence of loops of magnetization currents across the thickness of the SC. On the other hand, as the experimental measurements have shown negligible losses in the SS304 (stainless-steel) former, the electrical resistivity of it and the outer domain have been assumed as the one of `air' as it is custom in the electromagnetic modelling of superconductors at LN2 conditions. Thus, for the meshing of sample B, i.e., the one-layer cable, our simulations have considered a COMSOL physics-controlled mesh with a total of 10832 mesh vertices for 61283 Lagrange-shaped quadratic tetrahedrons, 9748 triangles, 1413 edge elements, and 92 vertex elements, with an average element quality of 0.55 in a mesh volume of $\sim1.63\times10^{-5}~m^{3}$. Likewise, for the two-layer cable (Sample C), the physics-controlled meshing function of COMSOL has render to a satisfactory model with 17396 mesh vertices for 98828 Lagrange-shaped quadratic tetrahedrons, 19210 triangles, 2818 edge elements, and 150 vertex elements, with an average element quantity of 0.5. Then, the alternating magnetic field is applied as a boundary condition in the $y-$direction of the outermost domain, with amplitudes ranging from $10-90$~mT. The model assumes the constant $J_{c}$ dependence and the $E(J)$ relation is defined by the conventional $E-J$ power law with $n$ exponent 30.5~\cite{Ruiz2019MDPI}.

\section{Results and discussion}
\label{Sec.3}

In order to disclose the most relevant electromagnetic quantities at a local and global level, this section focuses on the current distribution profiles inside the superconducting tapes and the overall AC losses of the CORC\textsuperscript{\textregistered} cables. The current path of the magnetization currents for the monolayer cable is shown at the left hand of Fig.~\ref{Fig_2}, it measured at the first peak of a 20~mT applied alternating magnetic field along the $y-$axis. The reader is encouraged to download and zoom this figure to appreciate its resolution. A top-view of the monolayer and bilayer cables is shown for their easy comparison with the conventional 2D results predicted by Bean-like models. Therefore, this reveals that the main direction of the magnetization currents is along the $z-$axis, i.e., closing the current loops across the edges of the tilted tapes. This proves that the classical predictions by Bean and the Critical State Model~\cite{Ruiz2009bPRB}, which have shown that the magnetization currents follow the minimum energy path in HTS bulks~\cite{Ruiz2012APL} and coated conductors~\cite{Ruiz2019MDPI}, are also seen in CORC\textsuperscript{\textregistered}  cables, i.e., with the loops of magnetization current mainly flowing in perpendicular direction to the applied magnetic field. This implies that in the case of the CORC\textsuperscript{\textregistered} cable, the outer surface of any HTS tape can show positive (light) or negative (dark) patterns of current density along the right and left-hand segments of the cable length, this depending on what direction the current must flow to screen the magnetic field by Ampere's law. Likewise, the opposite surface of the corresponding segment must show opposite current density profiles to screen the magnetic field from the inside of the superconducting layer. Then, as shown in the right-hand side of Fig.~\ref{Fig_2}, a similar electromagnetic response is seen within the six tapes of the bilayer CORC\textsuperscript{\textregistered} cable, i.e., obeying the same physical principles. The reason for this is twofold, as on the one hand the tapes are electrically insulated with Kapton tape (30~\micron thick) and, on the other hand the gaps between the three HTS tapes at the outer layer allows for the magnetic field to directly interact with the three HTS tapes at the inner layer. This 60~$\mu$m spacing between the tapes therefore causes the occurrence of magnetization currents across all HTS layers of the CORC\textsuperscript{\textregistered} cable.
      
\begin{figure}
\begin{center}
{\includegraphics[width=0.9\textwidth]{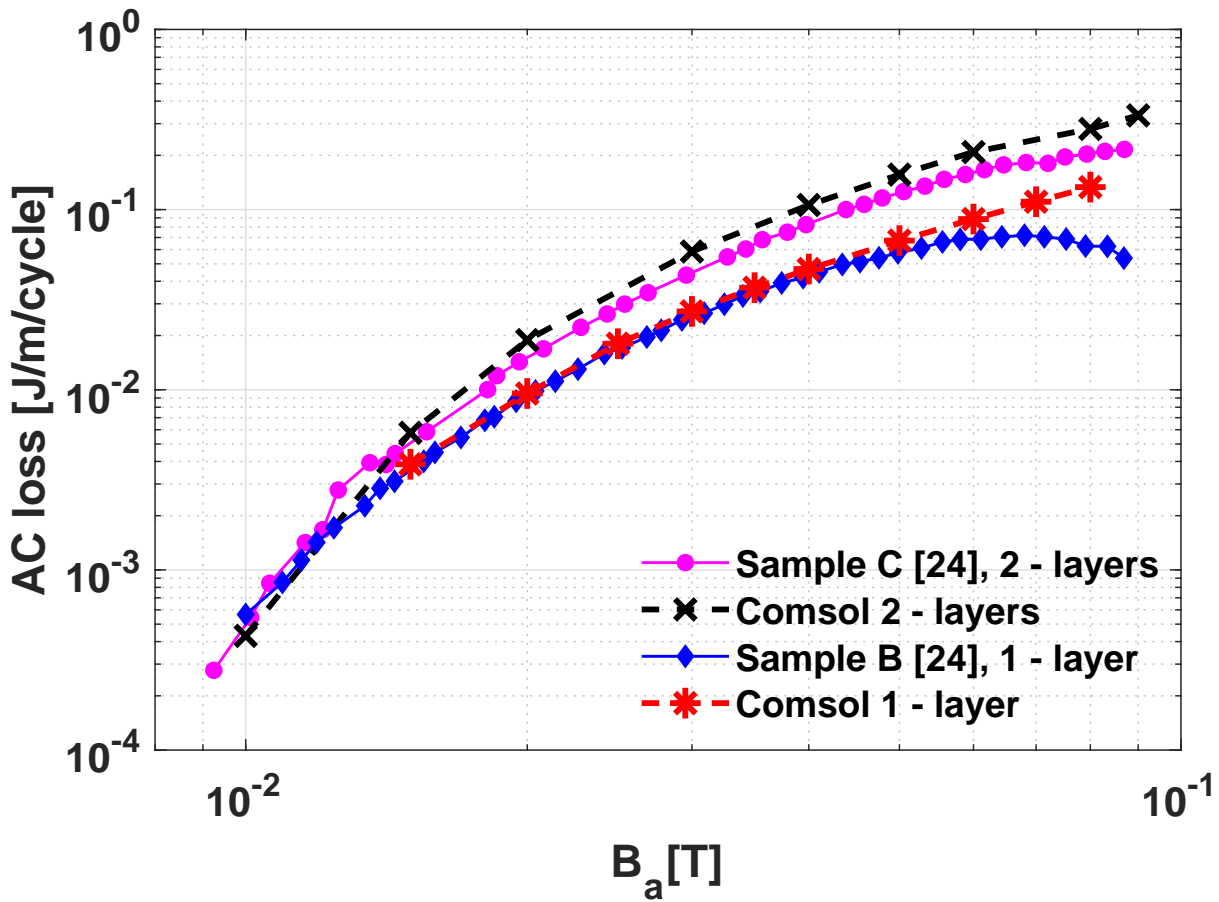}}
\caption{\label{Fig_3} Comparison between the experimental~\cite{Majoros2014} and computationally assessed AC losses for the CORC\textsuperscript{\textregistered} samples illustrated in Figs.~\ref{Fig_1}~\&~\ref{Fig_2}.
}
\end{center}
\end{figure}

Finally, being the AC losses one of the quantities of mayor practical interest, in Fig.~\ref{Fig_3} we compare our numerical results with the experimental measurements reported at~\cite{Majoros2014}. It shows a remarkable agreement for all the amplitudes of the magnetic field applied to the bilayer CORC\textsuperscript{\textregistered} cable (top-two curves), as well as with the monolayer cable up to 60~mT. In this latter case, possible heating effects have been reported, leading therefore to a reduced the critical current density of the HTS tapes and therefore to an apparent reduction of the AC losses in the coated conductors~\cite{Majoros2014}. Thus, our numerical model serves as the correct base for the estimation of AC losses in the CORC\textsuperscript{\textregistered} monolayer and bilayer cables designed by Advanced Conductor Technologies, as per their technical features reported in~\cite{Majoros2014}. This can be used as a benchmark model for an accurate and truly 3D modelling of CORC\textsuperscript{\textregistered} cables within the H-formulation.

%
\section{Conclusion}
\label{Sec.4}

In this article, we have shown and validated a full 3D FEM model within the so-called $H$-formulation as a study tool for the electromagnetic modelling and understanding of CORC\textsuperscript{\textregistered} cables, and their AC losses. The experimentally measured monolayer and bilayer cables of 3 and 6 tapes, respectively, have been modelled with the original manufacturer dimensions and physical properties introduced in COMSOL Multiphysics. Local and global electromagnetic quantities such as the three-dimensional distribution of current densities along and across the HTS tapes, and the overall AC losses of the CORC\textsuperscript{\textregistered} cable have been disclosed. Our results allow to elucidate the actual paths of the magnetization currents in helical arrangements of superconducting tapes, and how these resemble the classical Bean-like and critical state model observations in bulk and other coated conductor arrangements, without the need to recurring to artificial surface currents that occur in 3D to 2D mathematical reductions of the general PDE system, when for instance the thickness of the HTS tapes is reduced to an infinitely thin film approach. Moreover, good agreement with the experimentally measured AC losses for all magnetic fields applied to the bilayer CORC\textsuperscript{\textregistered} cable has been reported, what has allowed as to complete the curve of AC-losses for the monolayer cable at high magnetic fields, which as reported in \cite{Majoros2014} suffered of unintended heating effects. Thus, with computing times no longer than 20h for the entire hysteretic process of a bilayer 6-tapes CORC\textsuperscript{\textregistered} at 80~mT in a standard PC of 3.6GhZ clock speed and 32Gb RAM, the model here presented can be used with confidence as a benchmark for the modelling of CORC\textsuperscript{\textregistered} cables. The computing time can be greatly reduced by either considering larger time stepping, reduced meshing, reduced field, or enhanced computational power, but differences in the same order of magnitude for the computing time are of no importance as it is the physical validness of what must prevail.

\balance 
\vspace*{0.5cm}
\bibliographystyle{IEEEtran}
\bibliography{References_Ruiz_Group}

\end{document}